\documentclass[conference]{IEEEtran}
\usepackage{amsmath}
\usepackage{amssymb}
\usepackage{amsfonts}
\usepackage{graphicx}
\usepackage{float}
\usepackage[table, dvipsnames]{xcolor}
\usepackage{adjustbox}
\usepackage{booktabs}
\usepackage[short]{optidef}
\usepackage{multirow}
\usepackage{comment}

\usepackage{algorithm}
\usepackage{algpseudocode}
\usepackage{caption}
\usepackage{subcaption}

\usepackage{color,soul}

\DeclareRobustCommand*{\IEEEauthorrefmark}[1]{%
  \raisebox{0pt}[0pt][0pt]{\textsuperscript{\footnotesize #1}}%
}

\makeatletter
\newcommand*\titleheader[1]{\gdef\@titleheader{#1}}
\AtBeginDocument{%
  \let\st@red@title\@title
  \def\@title{%
    \bgroup\normalfont\large\centering\@titleheader\par\egroup
    \vskip.2em\st@red@title}
}
\makeatother

\title{Deep Reinforcement Learning-based Energy Efficiency Optimization For Flying LoRa Gateways}
\titleheader{2023 IEEE International Conference on Communications (ICC) // Rome, Italy}

\author{\IEEEauthorblockN{Mohammed Jouhari\IEEEauthorrefmark{1}, Khalil Ibrahimi\IEEEauthorrefmark{2}, Jalel Ben Othman\IEEEauthorrefmark{3} \IEEEauthorrefmark{4} \IEEEauthorrefmark{5} and El Mehdi Amhoud\IEEEauthorrefmark{1},
}
\IEEEauthorblockA{\IEEEauthorrefmark{1}School of Computer Science, Mohammed VI Polytechnic University, Ben Guerir, Morocco \\
\IEEEauthorrefmark{2}Ibn Tofail University, Faculty of Sciences, Laboratory of Research in Informatics (LaRI), Kenitra, Morocco\\
\IEEEauthorrefmark{3}CNRS, CentraleSupélec, Laboratoire des signaux et systémes, Université Paris-Saclay, Gif-sur-Yvette, France\\
\IEEEauthorrefmark{4}University of Sorbonne Paris North, Villetaneuse, France \\
\IEEEauthorrefmark{5}College of Technological Innovation, Zayed University, Abu Dhabi, UAE
} 
 mohammed.jouhari@um6p.ma, ibrahimi.khalil@uit.ac.ma, jalel.benothman@centralesupelec.fr, elmehdi.amhoud@um6p.ma
 \vspace{-4mm}}

\begin{document}



\maketitle
\begin{abstract}
A resource-constrained unmanned aerial vehicle (UAV) can be used as a flying LoRa gateway (GW) to move inside the target area for efficient data collection and LoRa resource management. In this work, we propose deep reinforcement learning (DRL) to optimize the energy efficiency (EE) in wireless LoRa networks composed of LoRa end devices (EDs) and a flying GW to extend the network lifetime. The trained DRL agent can efficiently allocate the spreading factors (SFs) and transmission powers (TPs) to EDs while considering the air-to-ground wireless link and the availability of SFs. In addition, we allow the flying GW to adjust its optimal policy onboard and perform online resource allocation. This is accomplished through retraining the DRL agent using reduced action space. Simulation results demonstrate that our proposed DRL-based online resource allocation scheme can achieve higher EE in LoRa networks over three benchmark schemes.
\end{abstract}

\IEEEpeerreviewmaketitle
\section{Introduction}
Long-range (LoRa) is one of the most popular low-power wide-area-network (LPWAN) protocols due to its easy deployment and flexible management as well as its open protocol stack. As a physical layer technology, LoRa adopts chirp spread spectrum (CSS) techniques to propagate narrowband signals over a specific channel bandwidth \cite{9357890}. The signal could therefore travel further while consuming less power, enabling the connection of thousands of devices with long battery lives. Based on different spreading factors (SFs), multiple LoRa end devices (EDs) can simultaneously share the same time slot and frequency channel. Hence, various signal-to-noise ratio (SNR) limits are required for each used SF, resulting in different data rate transmission and transmission range \cite{8647416}. For efficient LoRa communication, a decision on LoRa PHY layer parameters should be taken to select the adequate SF, transmission power (TP), and channel frequency depending on the wireless channel condition and the distance between the EDs and the gateway (GW). Thus an intelligent and proactive PHY LoRa parameters adjustment is needed owing to the time-variance of data traffic load, and channel condition \cite{9000517}.

The next generation of Internet of Things (IoT) networks is expected to benefit significantly from this technology \cite{etiabi2021huber, etiabi2022spreading}. However, this type of system generally involves using a large number of resource-constrained EDs to provide reliable data delivery to potentially critical applications, making LoRa resource allocation a complicated task \cite{jouhari2022survey, jouhari2017implementation}. Various works focused on LoRa resource management problems under different LoRa network scenarios and assumptions. To cope with concurrent transmission and interference, a cooperative SF assignment scheme was designed in \cite{9226528} for a multi-operator LoRaWAN deployment scenario. Game theory and gradient ascent-based iterative algorithms were used to solve the SF assignment optimization. Authors in \cite{9139964} suggested a hybrid LoRa-based low-power mesh network where EDs use both sub-GHz long-range radio and 2.4 GHz short-range radio. Furthermore, customized TDMA and ANT protocols were considered for efficient data collection in dense LoRa wireless networks. In contrast, authors in \cite{8884111} allocate LoRa PHY parameters in time slots for data transmission to alleviate LoRaWAN scalability issues by reducing collisions and grouping acknowledgments.

Moreover, deep reinforcement learning (DRL) was suggested in \cite{9530755} to efficiently manage the grid power consumption of GWs in hybrid LoRa Networks. Authors in \cite{10.1186/s13638-020-01783-5} combined centralized RL with distributed multi-agent RL to improve throughput and reduce the energy consumption of EDs. The authors in \cite{9802526} suggested freezing some layers when retraining an artificial neural network model on real data. This technique is intended to optimize the energy efficiency of LoRa systems by assigning the best power transmission to each device. In this work, we consider a LoRa network composed of EDs that sense data from the target region, and flying GW forwards collected data toward a LoRa server to establish the communication between the EDs and the server. The main contributions of this work are summarized as follows: 
\begin{itemize}
    \item We formulate the problem of SF and TP allocation as one optimization problem to maximize the energy efficiency (EE) in the LoRa network considering EDs position and air-to-ground (A2G) channel.
    \item We adopt a DRL agent to solve the EE optimization problem using the proximal policy optimization (PPO) algorithm to handle stochastic policies and cope with the high computation requirement of traditional RL.
    \item We design an efficient MDP environment with states, action space, and reward function that is adequate to the considered LoRa system for a stable learning of the DRL agent.
    \item We enable online resource management for moving flying GW by allowing fast policy adaptation onboard using an adequate action space reduction scheme (ASR).
\end{itemize}

This paper is outlined as follows: Section \ref{SystemModel} describes the system model and the problem formulation. DRL-based resource allocation scheme is detailed in Section \ref{DRL}, while Section \ref{sec:EvaluationResults} evaluates the performance of our proposed solutions. Finally, the paper is concluded in Section. \ref{sec:EvaluationResults}.
\section{System Model And Problem Formulation}\label{SystemModel}
\begin{figure}[t]
\vspace{1mm}
    \centering
    \includegraphics[width=0.98\linewidth]{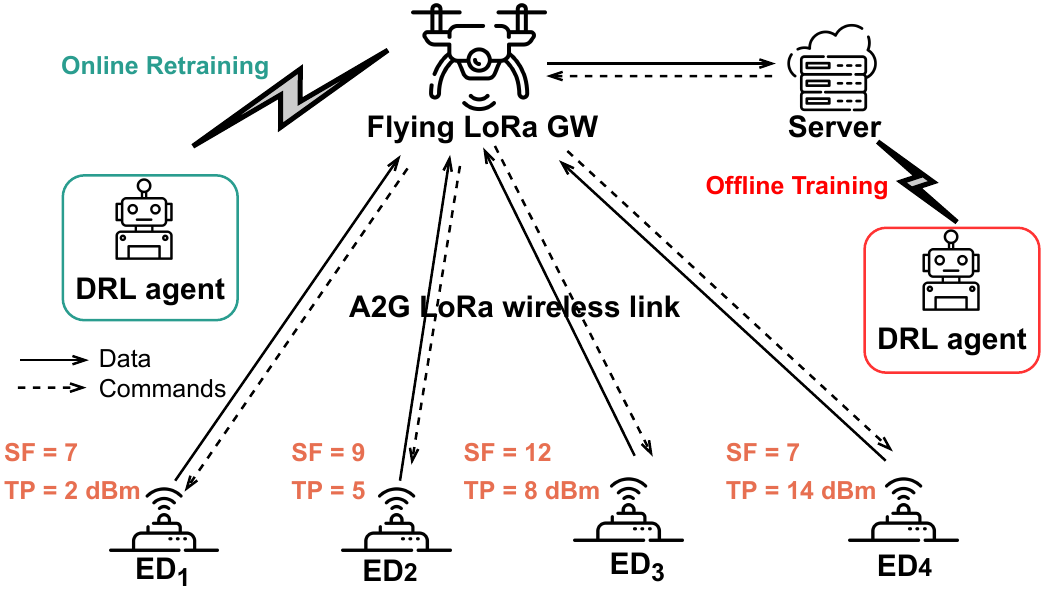}
    \caption{LoRa EDs transmit collected data to the flying GW using their allocated SF and TP}
    \vspace{-4mm}
    \label{fig:SystemModel}
\end{figure}

In this work, we consider a LoRa network composed of a set of $\mathcal{N}$ EDs and a single unmanned aerial vehicle (UAV) equipped with a LoRa GW to establish communication between EDs and the server (Fig. \ref{fig:SystemModel}). The UAV flies at a fixed altitude and changes its horizontal position periodically depending on the requirements of the mission and the events occurring in the monitored area. We characterize the A2G link by its path loss to associate the distance between the EDs and the GW with the link quality and LoRa specifications. This will help to identify the optimal TP required by the EDs to efficiently communicate their collected data to the GW. The path loss is composed of two elements $ l^{los}(i,g)$ and $ l^{nlos}(i,g)$ which are respectively the line-of-sight (LoS) and non-line of sight (nLoS) path loss and are given as follow:
\begin{equation}
    l^{los}(i,g) = l^{fs}(d_r) + 10\beta_{los}\log_{10}d_{i,g} + \chi_{\sigma_{los}},
\end{equation}
\begin{equation}
    l^{nlos}(i,g) = l^{fs}(d_r) + 10\beta_{nlos}\log_{10}d_{i,g} + \chi_{\sigma_{nlos}},
\end{equation}
where $\beta$ is the path loss exponent.
$\chi_{\sigma_{los}}$ and $\chi_{\sigma_{nlos}}$ are the shadowing random variables which are, respectively, characterized as the Gaussian random variables with zero mean and $\sigma_{los}$, $\sigma_{nlos}$ standard deviations.
$l^{fs}$ represent the free-space path loss in the reference distance $d_r$ which is calculated in dB by $l^{fs}(d_r) = 20\log_{10}(\frac{d_rf4\pi}{c})$,  
where $f$ is the carrier frequency in Hz, $c$ is the light speed, and $d_r$ the reference distance of the free space path loss. Since we consider communication between an UAV and multiple EDs anchored in the ground we assume that we have $d \geqslant c/f $ in such a way both antennas are far away from each other. Then according to the actual location of the UAV especially its flying altitude that causes occlusion, the LoS probability can be calculated as follows:
\begin{equation}\label{Equ:plos}
    P^{los}(\hat{\theta}) = \frac{1}{1 + \alpha e^{-\lambda(\hat{\theta} - \alpha)}},
\end{equation}
where $\alpha$ and $\lambda$ are the Sigmoid function parameters, while $\hat{\theta}$ refers to the elevation angle of the GW given by $\mathmbox{\hat{\theta}=\sin^{-1}(h_g/d_{i,g})}$. The probability value of nLoS can be obtained by $P^{nlos}(\hat{\theta}) = 1 - P^{los}(\hat{\theta})$. As results, the overall A2G path loss is calculated using both $l^{nlos}(i,g)$, $l^{nlos}(i,g)$:
\begin{equation}
    l^{a2g} = P^{los}(\hat{\theta})l^{los}(i,g) + P^{nlos}(\hat{\theta})l^{nlos}(i,g).
\end{equation}
The received signal strength indicator (RSSI) at the ED from the transmitting UAV which is our GW allows us to determine the quality of the LoRa communication link and to adjust the receiver sensitivity through tuning LoRa physical parameters. This relationship is defined based on the distance of the UAV and the nature of the communication environment that influences the direct visibility between the ED and the UAV. As described previously, the A2G path loss is composed of the LoS and nLoS parts related to the UAV position and altitude. Thus, the A2G path loss can also be obtained by substituting the RSSI value from the TP of the GW ($p^{tr}_{gw}$) as following
\begin{equation}\label{pl_rssi}
    l^{a2g}(d_{i,g}, \hat{\theta}) = p^{tr}_{gw} - RSSI.
\end{equation}
The RSSI value obtained from Eq. (\ref{pl_rssi}) in the path loss function and the TP of the GW should be greater than the receiver sensitivity to correctly decode the received information. In another way, the maximum distance where the received signal is decodable is obtained by substituting the sensitivity $\zeta_{n,m}$ instead of the RSSI in Eq. (\ref{pl_rssi}). 
To link the network topology with LoRa PHY parameters we consider the signal-to-noise ratio between the GW and the EDs using the SF $m$ defined as $\mathrm{\rho}^{a2g}_{n,m} = \frac{p_{n,m}}{10^{0.1l^{a2g}(d_{n}, \hat{\theta})}.\sigma^2}$. $p_{n,m}$ is the TP of EDs $n$ at SF $m$ while $\sigma^2$ refers to the Gaussian noise power. Hence, the signal-to-interference- plus-noise ratio $\Upsilon$ is given by:
\begin{equation}
    \mathrm{\Upsilon}_{n,m} = \frac{\mathrm{\rho}^{a2g}_{n,m}}{\sum\limits_{k \in \mathcal{N}_{- n}} \psi_{k,m}\mathrm{\rho}^{a2g}_{k,m} + 1},
\end{equation}
Where $\mathcal{N}_{- n} = \mathcal{N} \backslash \{n\}$ and $\psi_{k,m}$ is a binary variable that indicates whether the ED k uses the SF m from the set of SF $\mathcal{M}$.
The channel capacity between ED $n$ and the GW 
can be calculated by $\Gamma_{n,m} = \mathrm{BW}\log_2(1 + \mathrm{\Upsilon}_{n,m}),$ where BW is the bandwidth.
We define the energy efficiency (EE) of the system by:
\begin{equation}
    \xi_{EE} = \frac{\Gamma}{P_T + P_c},
\end{equation}
where $P_c$, $P_T = \sum\limits_{n=1} p_{n,m}$ and $\Gamma$ are respectively the circuit consumption power, the TP and the sum rate of the system over all the logical channels given as $\Gamma = \sum\limits_{n=1} \Gamma_{n,m}.$

Our objective is to maximize the EE of the LoRa network by allocating the best SF and TP depending on the ED position and the A2G channel quality. For this we formulate the following optimization problem:
\begin{maxi!}{\substack{\psi_{n,m} \in \Psi \\p_{n,m} \in P}}{\xi_{EE}}{\label{equ:p1}}{\text{(P1)}}
\addConstraint{0\leq p_{n,m}\leq p_{max}}{}
\addConstraint{\psi_{n,m} \in \{0, 1\} \forall n,m}{}
\addConstraint{\sum\limits_{m=1}\psi_{n,m} \leq 1,\forall n}{}
\addConstraint{\sum\limits_{n=1}\psi_{n,m} \leq \varrho_{max},\forall m}{}
\addConstraint{\zeta_{n,m} \leq RSSI_{n,m},\forall n,m}{},
\end{maxi!}
where $\Psi = \{\psi_{n,m}|m \in \mathcal{M}, n \in \mathcal{N}\}$, $P = \{p_{n,m}| n \in \mathcal{N}, m \in \mathcal{M}\}$ denote the optimization variable sets of EDs clustering based on SF assignment and the TP allocation. The optimization problem (P1) is constrained by the lower and the upper bounds of the TP available for each ED (\ref{equ:p1}b) and the binary allocation variable $\psi_{n,m} \in \{0, 1\}$ (\ref{equ:p1}c). Constraint (\ref{equ:p1}d) implies that each ED can adopt at most one SF. To cope with the limited available SF multiple EDs can be associated with the same SF (\ref{equ:p1}e) \cite{8647416}. The constraint (\ref{equ:p1}f) guarantees that the received signal power is detectable at the ED.

\section{DRL-based resource allocation}\label{DRL}
The problem (P1) is known to be NP-hard, which makes the efficient allocation of TP and SF more complicated and costly in terms of computation resources and time. Moreover, solving this problem requires full knowledge of the network topology and the assigned SF in the future time step to deal with co-SF interference. This is not practical in a real-life scenario. DRL techniques have shown excellent performance as a tool to solve such online problems known by their complex and dynamic spaces. It takes advantage of the interaction with the environmental parameters (e.g., A2G link budget, energy costs) to learn the most efficient policy from the statistical distribution of these features and the system's current state using its insights of the future learned through experiences. Hence the DRL technique seems to be an adequate solution to solve problem (P1) while reducing its complexity by making suboptimal decisions following the learned policy.

Previously, we described the optimization (P1) to efficiently maximize the system's energy efficiency while considering the A2G link budget, co-SF interferences, and the network topology. The optimization is performed by assigning the adequate SF and TP that maximize the energy efficiency. Accordingly, we designed a DRL agent to solve the problem mentioned above. In particular, at each time step of the resource allocation process, the DRL agent decides on the SF and TP to allocate for each EDs regarding the system's current state. The DRL agent ultimately aims to maximize the channel throughput by reducing co-SF interferences between EDs while reducing the TP regarding the distance between the EDs and the UAV-based GW. During the training process, the agent receives rewards for each allocation made until convergence owing to the optimal policy. The convergence of the DRL agent is achieved when its learning curve gets flat and stops increasing. To implement the solution, we abstracted the problem into a Markov Decision Process (MDP) framework of the four-tuple ($\mathcal{S}$, $\mathcal{A}$, $R$, $\mathcal{P}_r$). Such that, $\mathcal{S}$ depicts the state set of the system, $\mathcal{A}$ the set of actions, $R$ is the immediate reward that can be received for each action, and $\mathcal{P}_r$ the probability of transition. 

\begin{figure}[t]
    \centering
    \includegraphics[width=0.98\linewidth]{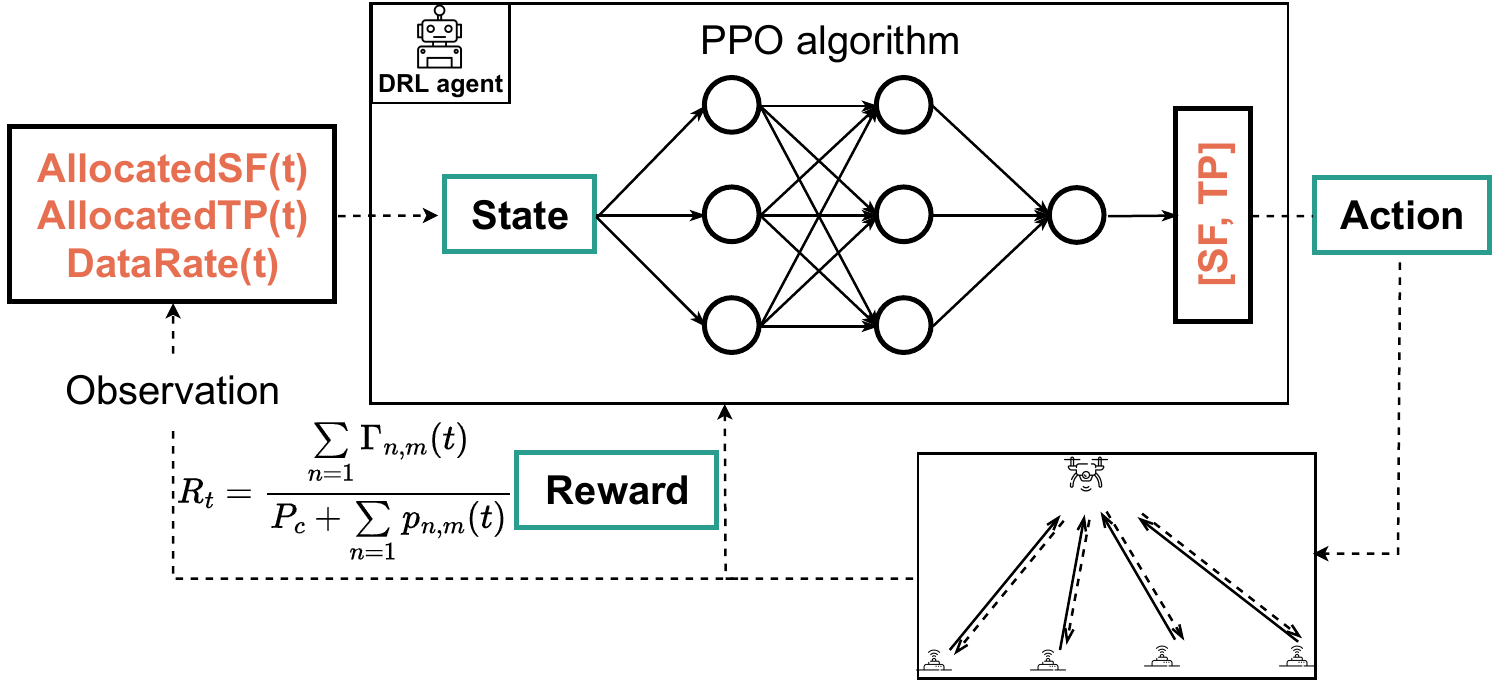}
    \caption{DRL architecture}
    \label{fig:drlarchi}
    \vspace{-5mm}
\end{figure}
\subsubsection{MDP environment}
The DRL agent interacts with the LoRa network (as shown by Fig. \ref{fig:drlarchi}) presented as an MDP environment, designed to generate a reward $R_t$ for each agent action $\mathcal{A}_t$ at time-step t that impacts the next state on the system $\mathcal{S}_{t+1}$. An episode consists of a consecutive sequence of some time steps. Throughout multiple episodes, the agent is trained by experiencing various scenarios to learn the optimal policy from its previous actions and the received rewards. At each time step, the agent decides on assigning the TP and SF for single LoRa ED as described in the problem (P1). Episodes are independent, meaning that the cumulative reward is initiated to 0 at the beginning of each episode. Moreover, a new network topology is considered in each episode. Hence, the episode length corresponds to the number of active LoRa EDs in the network, which should remain the same during the learning process. Since the agent does not have an overall overview of the environment, thus the policy is obtained through learning from its interaction with the environment.
\subsubsection{State and action spaces}
We set the state space as $\mathcal{S} = \{ AllocatedSF(t), AllocatedTP(t), DataRate(t) \}$ where $AllocatedSF(t)$ and $AllocatedTP(t)$ are respectively the matrices of allocated SF and TP at time step $t$. $DataRate(t)$ is a matrix of the current data rate between the EDs and the GW after assigning resources to the ED corresponding to the time step $t$. The DRL agent takes actions from the action space $\mathcal{A}_t$ based on the current state of the system and its policy $\Pi$. This action consists of allocating a combination of the SF and the TP which can be expressed as $\mathcal{A}_t = \{ m_t, p_t\}$ such as $m_t \in \{ 7, 8, 9, 10, 11, 12\}$ and $p_t \in \{2, 5, 8, 11, 14\}$. LoRa ED relevant to the current step is served through the allocated resources once the decision is made. As each step is completed, the allocated SF and TP matrices and the data rate map are updated based on the current decision. The new matrices are then used as input for the next step.
\subsubsection{Immediate reward function R}
We formulate the immediate reward function to match the objective of our optimization problem, aiming to maximize the energy efficiency of all EDs in the network while respecting the availability of SF and the TP to meet the sensitivity requirement. Specifically, when the DRL agent observes the state of the system $\mathcal{S}_t$ an action $\mathcal{A}_t$ should be taken that meets the two constraints:
\begin{equation}\label{equ:constraints}
  \begin{cases}
    C1:  & \sum\limits_{i=1}^{i=t}\mathcal{A}_{i}(0) \leq  6, \, \text{Constraint \, (19e)}\\ 
    C2:  & \zeta(\mathcal{A}_t(0)) \leq RSSI, \, \text{Constraint \, (19f).} 
  \end{cases}
\end{equation}
(C1) indicates that a SF can not be assigned to more than 6 EDs which depicts the channel capacity for handling the co-SF interference and matches the constraint (19e). (C2) means that the transmitted signal strength using the TP $\mathcal{A}_t(1)$ should be more significant than the receiver sensitivity using the SF $\mathcal{A}_t(0)$. Therefore, the immediate reward is formulated as follows:
\begin{equation}\label{Equ:reward}
    R_t = \frac{ \sum\limits_{n=1} \Gamma_{n,m}(t)}{P_c +  \sum\limits_{n=1} p_{n,m}(t)}.
\end{equation}
A null reward is provided when only one constraint is not respected. Otherwise, a positive reward calculated from Eq. (\ref{Equ:reward}) is provided to the agent, which helps to find the optimal policy by maximizing the cumulative rewards through consecutive time steps.
\subsubsection{DRL algorithm}
As described previously, we are working on the time-variant LoRa system with a dimensional set of actions that depends on the available SF and the TP level. Hence, using traditional RL methods and saving all experienced MDP tuples in a table can be challenging and requires high computation capability. To cope with this, we adopted the DRL technique based on DNN to solve our optimization problem. The considered DRL characterized by its dynamic state space is deployed using the proximal policy optimization (PPO) \cite{schulman2017proximal}, known for its high performance for stochastic policies which makes it an adequate solution for our problem.
\setlength{\textfloatsep}{6pt}
\setlength{\floatsep}{6pt}
\begin{algorithm}
\caption{PPO algorithm training process}\label{alg:cap}
\begin{algorithmic}[1]
\State - Randomly initialize $\theta$ and get $\Pi_\theta$
\State - Establish a sampling policy $\Pi_{\theta^*}$ such as $\theta^* \gets \theta$
\For{each episode e}
\State $\mathcal{S}_0 \gets$ NewTopology
\For{$t \in \mathcal{N}$}
    \State take $\mathcal{A}_t$ based on the policy $\Pi_{\theta^*}$
    \If{$\mathcal{A}_t$ respects (C1) and (C2)}
    \State Update $\mathcal{S}_t$ using the allocated resources
    \EndIf
\EndFor
    \If{Resources allocated to all EDs}
    \State Calculate the current reward $R_t$
    \State Observe the next state $\mathcal{S}_{t+1}$ and $R_t$
    \State Save $(\mathcal{S}_t, \mathcal{A}_t, R_t, \mathcal{S}_{t+1})$ in D
    \EndIf
    \State - Compute the estimator $\hat{E}_t = \sum_{i=0}^{\infty} (\gamma\chi)^i \delta_{t+i}$
    \State - Where $\delta_{t+i} = R_t + \gamma V(\mathcal{S}_{t+1}, \theta) - V(\mathcal{S}_t, \theta)$
    \State - Select mini-batch sample $(\mathcal{S}_t, \mathcal{A}_t, R_t, \mathcal{S}_{t+1})$ from D
    \State - Update $\theta$ based on the sampled data and maximizing: 
    \State $L^{CLIP}(\theta) = \mathbb{E}[p_t(\theta)\hat{E}_t, clip(p_t(\theta), 1-\epsilon, 1+\epsilon)\hat{E}_t]$
    \State - Where $p_t(\theta)$ is calculated as $p_t(\theta)= \frac{\Pi(\mathcal{A}_t\|\mathcal{S}_t,\theta)}{\Pi(\mathcal{A}_t\|\mathcal{S}_t,\theta^*)}$
    \State - Update DNN parameters: $\theta^* \gets \theta$
\EndFor
\end{algorithmic}
\end{algorithm}
The training process of the PPO algorithm to reach convergence is described in algorithm \ref{alg:cap}, including multiple steps with respect to our MDP model. We start the training process by initializing the DNN weights to get initial PPO policies used to generate samples for different episodes using $\theta^*$. At each episode, a new network topology is considered, which consists of the new position of EDs in the target area (line 4). An action $\mathcal{A}_t$ is taken based on the current policy $\Pi_{\theta^*}$. If the action $\mathcal{A}_t$ respects the constraints (C1) and (C2), the resources will be assigned to the concerned EDs, and the system state will be updated (lines 5-10). D is a replay memory where these samples are stored (lines 12-14). At the end of each episode, the policy gradient estimator $\hat{E}_t$ is calculated in lines 16-17. In line 18, samples are extracted from a random mini-batch from the experience memory D. The parameter $\chi$ is used to regulate the bias-variance trade-off. In contrast, $V(\mathcal{S}_t, \theta)$ is the state value describing the expected return of the system in state t using the parameters $\theta$. The objective function $L^{CLIP}(\theta)$ (line 20) is maximized for each batch sample to update the main policy $\Pi_\theta$. The main goal of the clip function is to keep the policy probability ratio in the range of (1-$\epsilon$) and (1+$\epsilon$). This mechanism used to learn from past experiences, known as experience replay, is vital to converge and stabilize learning over time. Finally, both policies are synchronized by replacing the oldest with the new one.
\begin{figure*}[t]
     \centering
     \begin{subfigure}[]{0.3\textwidth}
         \centering
         \includegraphics[width=\textwidth]{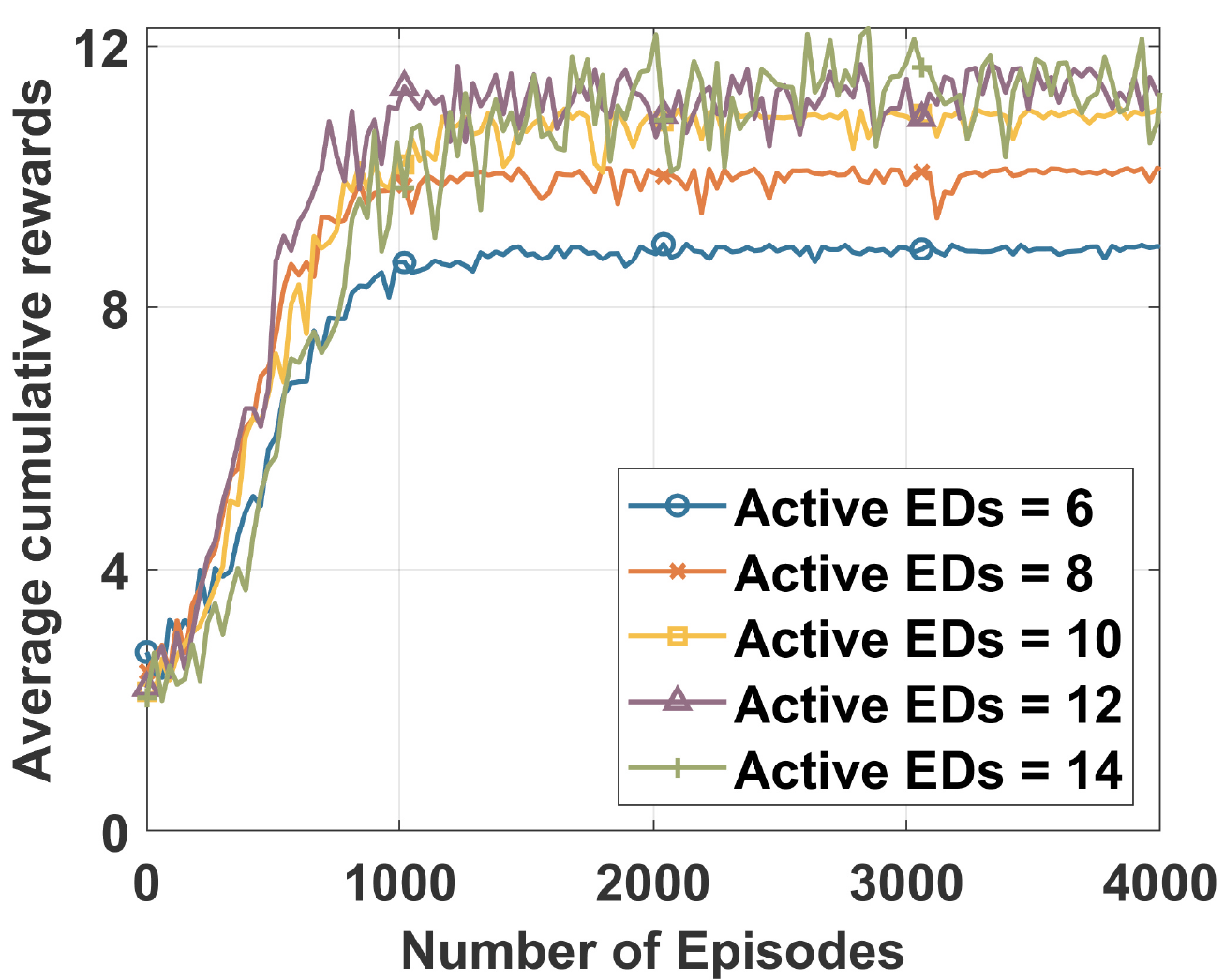}
         \caption{}
     \end{subfigure}
     \hfill
     \begin{subfigure}[]{0.3\textwidth}
         \centering
         \includegraphics[width=\textwidth]{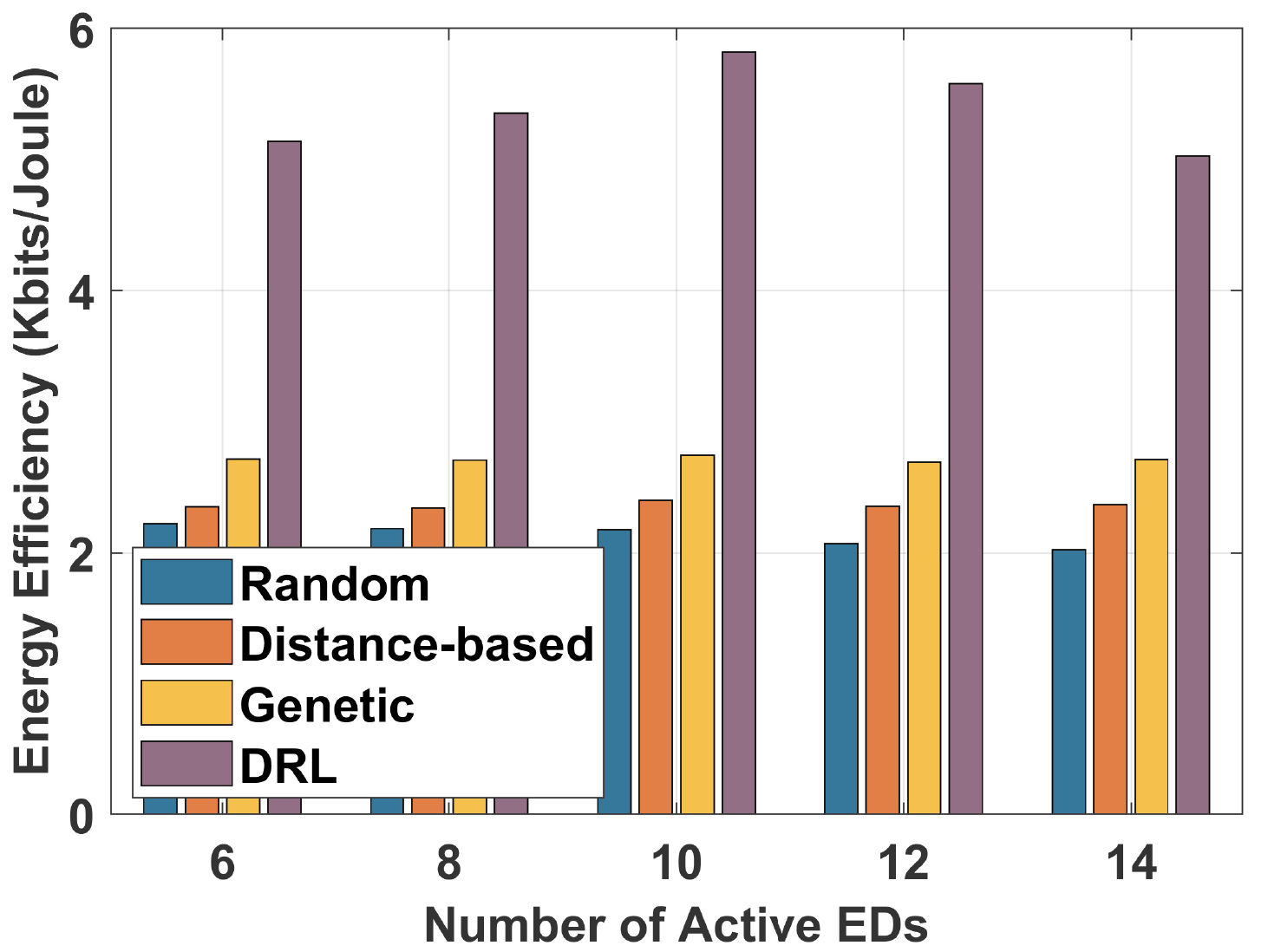}
         \caption{}
     \end{subfigure}
     \hfill
     \begin{subfigure}[]{0.3\textwidth}
         \centering
         \includegraphics[width=\textwidth]{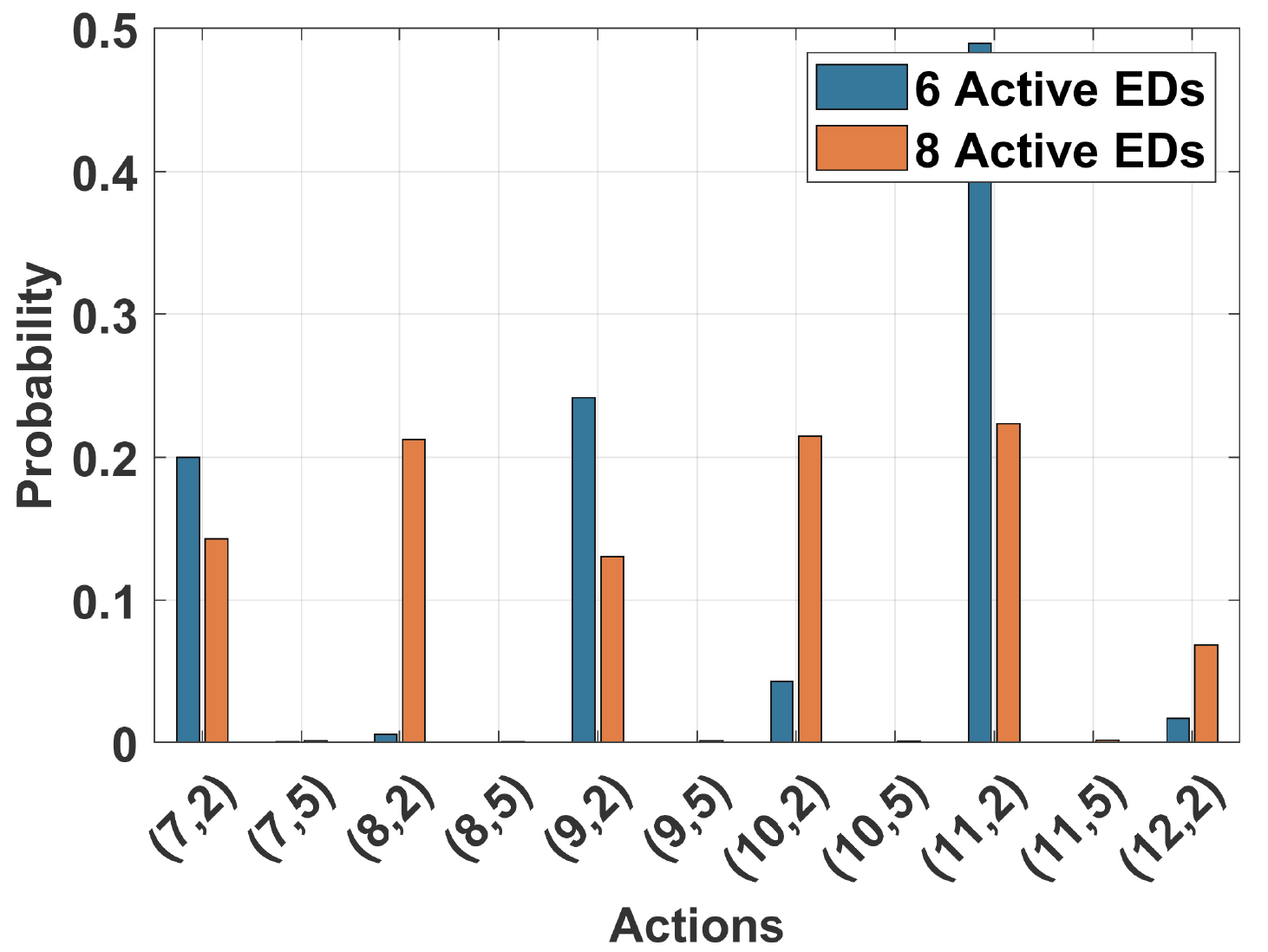}
         \caption{}
     \end{subfigure}
        \caption{Performance of DRL agent: (a) Training rewards for each network density (b) EE comparison with benchmarking schemes (c) Probability distribution of actions.}
        \label{fig:train_ee}
        \vspace{-3mm}
\end{figure*}
\section{Performance Evaluation}\label{sec:EvaluationResults}
In this section, we simulate our proposed model to evaluate its performances and compare it to the random, distance-based, and genetic algorithms \cite{reynders2017power}. The distance-based scheme allocates SF and TP  based on the distance of EDs from the flying GW. The EDs positions are generated randomly among the target area of $2 \,km^2$ at each episode in the training process. The flying GW is a UAV located at the center of this area with an altitude of 0.3 km and is used to collect data from active EDs using LoRa protocol. Parameters of the wireless link and the DRL agent used in this simulation to evaluate our optimization are fixed as follows, $\gamma = 0.99$, $\epsilon = 0.1$, $\chi = 0.01$. Also, we have considered the path loss exponent $\beta_{los}=2$ and $\beta_{nlos}=2.5$, the Gaussian random variable $\sigma_{los} = 5$ and $\sigma_{nlos} = 20$, f = 868 GHz, BW = 125 kHz.

Fig. \ref{fig:train_ee} (a) shows the convergence of the average cumulative reward of the DRL agent in the training process versus the number of episodes. The agent is trained through experiencing different network topologies of fixed UAV and variable EDs positions for multiple consecutive episodes to learn the optimal policy. As shown from the figure, the agent reached the convergence after 1200 episodes for different network densities and learned the best policy to allocate SF and TP while considering the system's energy efficiency and the A2G LoRa wireless link. Moreover, networks with low density obtain low rewards in the training process, which does not mean achieving low energy efficiency; it depends on the agent learning of the relation between variable EDs positions and the fixed flying GW. Fig. \ref{fig:train_ee} (b) illustrates the performance comparison of our proposed energy efficiency optimization solution using a DRL agent versus traditional algorithms such as the random resource allocation and distance based where the higher SF are assigned to further EDs. On the other hand, the key idea of the genetic algorithm is to assign different SFs and TP to different EDs such that constraints (C1) and (C2) are respected. This algorithm is configured with a population size of 200, elite size of 20 and mutation rate of 0.6, and 5000 generations. The proposed DRL-based optimization solution outperforms the considered traditional scheme in terms of energy efficiency due to the power of reinforcement learning in handling the considered LoRa system, including the A2G wireless link, EDs position, the available SFs, and the dimensional action space. In Fig. \ref{fig:train_ee} (c), we plot the probability distribution of actions for network densities of 6, and 8, knowing that as defined previously an action is a tuple of (SF, TP). In Fig. \ref{fig:train_ee} (c),  we plotted only the actions with a probability higher than 0.001. This will help us to identify the most taken actions by the agent or the preferred actions and reduce the action space dimension for future use, as we will show after. As it can be noticed from the figure, all the SF can be assigned to the EDs with different probabilities depending on their availability while respecting the constraint (C1) and the distance of EDs from the GW. However, due to the energy efficiency purpose, the agent favored the low TP level. Consequently, TP of 11 dBm and 14 dBm were not taken despite being available in the action space.  
\begin{figure}[ht]
     \centering
     \begin{subfigure}[]{0.24\textwidth}
         \centering
         \includegraphics[width=\textwidth]{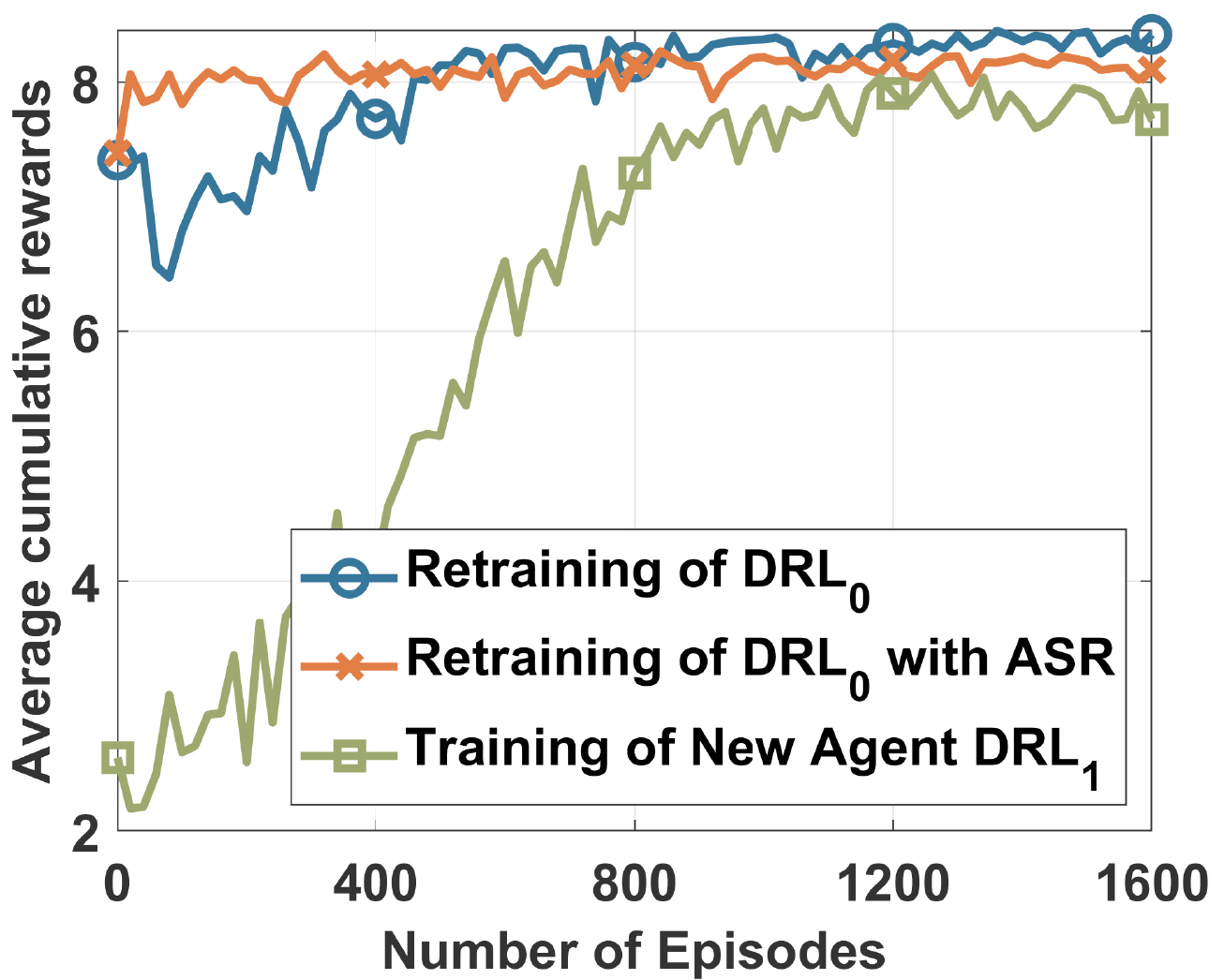}
         \caption{N = 6}
         \label{fig:N6}
     \end{subfigure}
     \hfill
     \begin{subfigure}[]{0.24\textwidth}
         \centering
         \includegraphics[width=\textwidth]{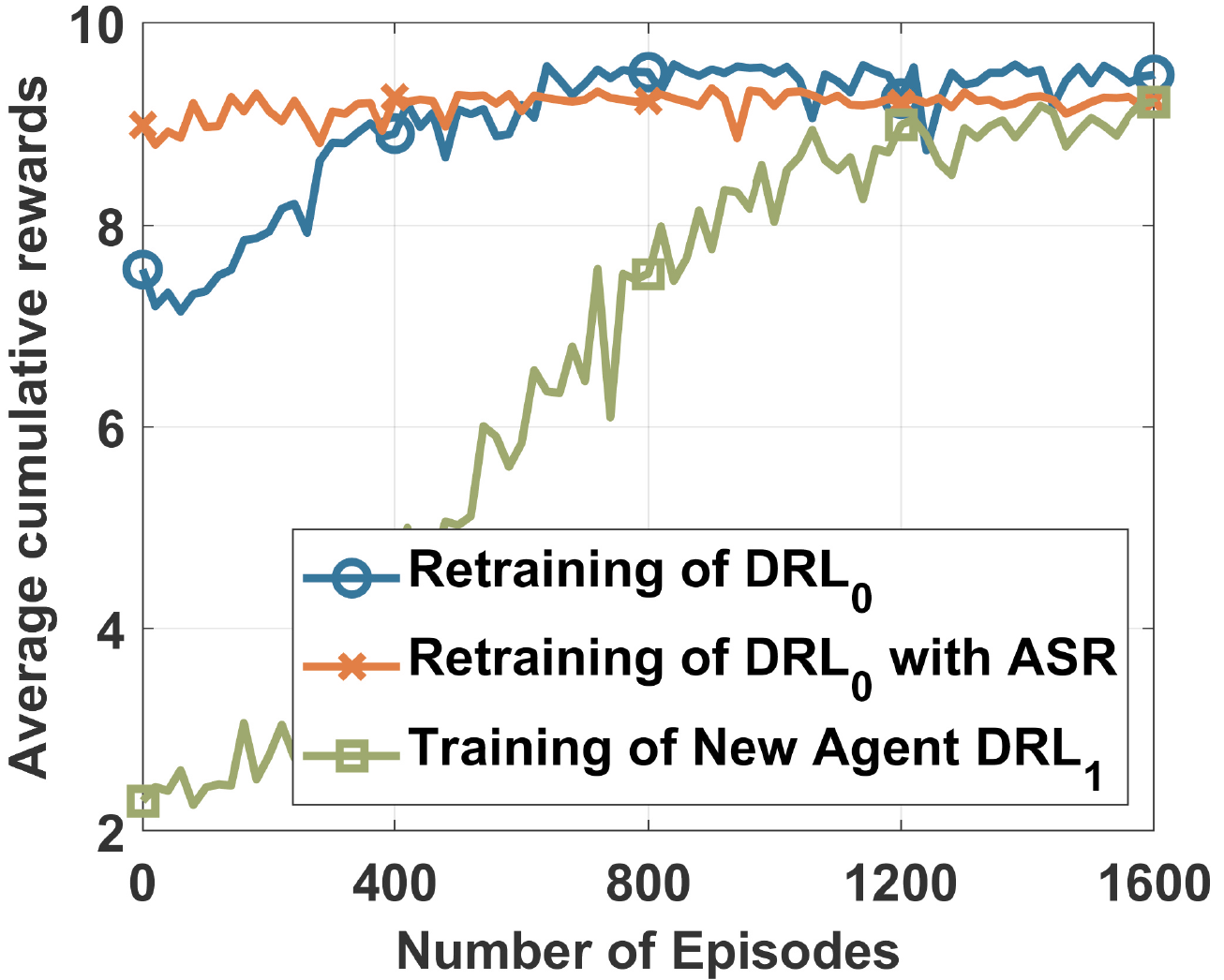}
         \caption{N = 8}
         \label{fig:N8}
     \end{subfigure}
        \caption{Training results of DRL on the new configuration}
        \label{fig:conv}
\end{figure}
\newline
We consider a new scenario where the flying GW moves from its initial position in the center of the target area to a new position in the corner. In Fig. \ref{fig:conv}, we show the outcome of the training process using different network configurations where the flying GW moves from its initial position 0 to the new position 1 using the pre-trained DRL$_0$ agent, which is trained on the network configuration 0 to infer in the network configuration 1 results in performance reduction of the system in terms of energy efficiency. To cope with this, a typical solution can be used by training a new agent using the new configuration, as shown by the green curve in Fig. \ref{fig:conv} (a). This process is generally performed offline, which requires time and computation resources to ensure stable system training and convergence. However, time and computation resources may not be available for flying GW to deal with the time-variance of event occurrence in the target area. Thus we suggest continuing the training of the pre-trained DRL$_0$ agent (blue curve Fig. \ref{fig:conv} (a)) to reduce both the convergence time and the required computation resources to allow an online policy adjustment onboard the UAV through a small training process. In addition, we used the results of Fig. \ref{fig:train_ee} (c) to reduce the action space of the agent by eliminating the actions of low probability, which helped further reduce the convergence time of the retraining process in the cost of the energy efficiency (red curve Fig. \ref{fig:conv} (a)). The experience is performed for networks composed of 6 and 8 active EDs (Fig. \ref{fig:conv} (b)) to show further the system's behavior against the action space reductions.
\begin{table*}[t]
    \centering
    \vspace{3mm}
    \renewcommand{\arraystretch}{1.1}
    \caption{Results summary of our proposed DRL-based energy efficiency optimization}
    \begin{tabular}{c | c c | c c | c c | c c | c c}
        \toprule
        \multirow{3}{*}{\textbf{Active EDs}} & \multicolumn{2}{c|}{\textbf{Scenario 0}} & \multicolumn{2}{c|}{\textbf{Scenario 1}} & \multicolumn{2}{c|}{\textbf{Scenario 1}} & \multicolumn{2}{c|}{\textbf{Scenario 1}} & \multicolumn{2}{c}{\textbf{Scenario 1}}\\
        & \multicolumn{2}{c|}{\textbf{Trained DRL$_0$}} & \multicolumn{2}{c|}{\textbf{Pretrained DRL$_0$}} & \multicolumn{2}{c|}{\textbf{Retrained DRL$_0$}} & \multicolumn{2}{c|}{\textbf{Retrained DRL$_0$ with ASR}} & \multicolumn{2}{c}{\textbf{Trained DRL$_1$}}\\
        \cline{2-11}
    & \textbf{EE} & \textbf{Accuracy} & \textbf{EE} & \textbf{Accuracy} & \textbf{EE} & \textbf{Accuracy} & \textbf{EE} & \textbf{Accuracy} & \textbf{EE} & \textbf{Accuracy} \\
    \bottomrule
    \textbf{6} & 5.14 & 100 & 5.05 & 100 & 5.26 & 100 & 5.12 & 99.7 & 5.07 & 100\\
    
    \textbf{8} & 5.35 & 99.8 & 4.95 & 91.1 & 5.49 & 100 & 5.19 & 98.1 & 5.32 & 100\\
    
    \textbf{10} & 5.82 & 99.9 & 5.66 & 87.9 & 5.85 & 100 & 5.73 & 94.4 & 5.79 & 98.83\\
    
    \textbf{12} & 6.02 & 88.3 & 5.77 & 79.5 & 5.87 & 95.5 & 5.72 & 77.9 & 6.19 & 97.29\\
    
    \textbf{14} & 6.1 & 93.7 & 5.15 & 77.7 & 5.92 & 99.1 & 5.95 & 84.7 & 6.23 & 94.23\\
    
    \bottomrule
    \end{tabular}
    \vspace{-6mm}
    \label{tab:infer_comp}
\end{table*}
Table. \ref{tab:infer_comp} summarizes the inference results of our proposed DRL-based energy efficiency optimization solution to deal with moving flying LoRa GW. The second column shows the test results of DRL$_0$ trained and tested using the network where the GW is located in the center of the area with different EDs positions. Energy efficiency and accuracy are both considered as performance metrics to evaluate the proposed solutions. EE depicts the average energy efficiency of all active EDs in the network, and the accuracy is the percentage of tests where the agent satisfied both constraints (C1) and (C2). Obtained results show that networks with lower density achieve high accuracy due to the max SF reuse in constraint (C1), which is easy to respect in networks with low density. Although the EE increases with the density due to the short distance between the EDs and the GW, which requires low TP to reach the destination conducting to higher energy efficiency. This trend is almost followed for any network configuration. The system performance degrades when using the pre-trained DRL$_0$ to allocate resources in network configuration 1. This is a normal result since the agent is only trained to learn to allocate resources for fixed GW and moving EDs. As shown in Table \ref{tab:infer_comp}, retraining the DRL$_0$ improves its network performance with the new GW position. However, to reduce the convergence time of the DRL$_0$ training on the new network configuration, we used the ASR method. The obtained results improve those of DRL$_0$ while being lower than retrained DRL$_0$ due to the set of actions removed from the action space which conduct the agent to take inadequate action for the energy efficiency of the system. 
The last column of the table shows that training a new DRL agent does not provide good results in terms of EE owing to the important convergence time required for the training process and the computation resource required for that which is not available in flying GW.
\vspace{-6mm}
\section{Conclusion}
In this work, we proposed a DRL-based resource allocation to maximize the energy efficiency in the LoRa network composed of  EDs deployed randomly in the target area and a single flying GW. This latter manages LoRa resources to avoid interference between EDs and allows efficient data transmission. Furthermore, the flying GW can change its position following the events occurring in the target area to increase the throughput of EDs in this region, reducing communication latency and improving the system's energy efficiency. In this work, we suggested using the PPO algorithm to solve the energy efficiency optimization by assigning SF and TP to EDs in the network while considering constraints on co-SF interference and the A2G LoRa wireless link. The obtained results outperformed the existing ones in terms of energy efficiency due to PPO's ability to deal with stochastic behavior and dimensionality. In addition, to enable a real-time decision of flying GW when changing its position, we proposed retraining the pre-trained DRL agent using ASR. In future works, we look forward to incorporating simultaneously two DRL agents, for LoRa resource allocation and for data collection optimization.
\vspace{-3mm}
\section*{Acknowledgement}
This work was sponsored by the Junior Faculty Development program under the UM6P – EPFL Excellence in Africa Initiative, and partly supported by ASPIRE, the technology program management pillar of Abu Dhabi’s Advanced Technology Research Council (ATRC), via the ASPIRE Visiting International Professorship program.
Project ASPIRE/ZU R22036 EU2105.
\vspace{-3mm}
\bibliographystyle{ieeetr}
\bibliography{library}

\begin{thebibliography}{10}

\bibitem{9357890}
P.~A. Frangoudis, C.~Tsigkanos, and S.~Dustdar, ``{Connectivity Technology
  Selection and Deployment Strategies for IoT Service Provision Over LPWAN},''
  {\em IEEE Internet Computing}, vol.~25, no.~1, 2021.

\bibitem{8647416}
B.~Su, Z.~Qin, and Q.~Ni, ``{Energy Efficient Resource Allocation for Uplink
  LoRa Networks},'' in {\em IEEE Global Communications Conference (GLOBECOM)},
  pp.~1--7, 2018.

\bibitem{9000517}
A.~Furtado, J.~Pacheco, and R.~Oliveira, ``{PHY/MAC Uplink Performance of LoRa
  Class A Networks},'' {\em IEEE Internet of Things Journal}, vol.~7, no.~7,
  pp.~6528--6538, 2020.

\bibitem{etiabi2021huber}
Y.~Etiabi, E.~M. Amhoud, and E.~Sabir, ``{Huber estimator and statistical
  bootstrap based light-weight localization for IoT systems},'' in {\em
  International Symposium on Ubiquitous Networking}, pp.~79--92, Springer,
  2021.

\bibitem{etiabi2022spreading}
Y.~Etiabi, M.~Jouhari, and E.~M. Amhoud, ``{Spreading Factor and RSSI for
  Localization in LoRa Networks: A Deep Reinforcement Learning Approach},''
  {\em arXiv preprint arXiv:2205.11428}, 2022.

\bibitem{jouhari2022survey}
M.~Jouhari, N.~Saeed, M.-S. Alouini, and E.~M. Amhoud, ``{A Survey on Scalable
  LoRaWAN for Massive IoT: Recent Advances, Potentials, and Challenges},'' {\em
  IEEE Communications Surveys \& Tutorials}, pp.~1--1, 2023.

\bibitem{jouhari2017implementation}
M.~Jouhari, K.~Ibrahimi, and M.~Benattou, ``Implementation of bit error rate
  model of 16-qam in aqua-sim simulator for underwater sensor networks,'' in
  {\em Advances in Ubiquitous Networking 2: Proceedings of the UNet’16 2},
  pp.~123--134, Springer, 2017.

\bibitem{9226528}
H.~Fawaz, K.~Khawam, S.~Lahoud, S.~Martin, and M.~E. Helou, ``{Cooperation for
  Spreading Factor Assignment in a Multioperator LoRaWAN Deployment},'' {\em
  IEEE Internet of Things Journal}, vol.~8, no.~7, 2021.

\bibitem{9139964}
X.~Jiang, H.~Zhang, E.~A. Barsallo~Yi, N.~Raghunathan, C.~Mousoulis,
  S.~Chaterji, D.~Peroulis, A.~Shakouri, and S.~Bagchi, ``{Hybrid Low-Power
  Wide-Area Mesh Network for IoT Applications},'' {\em IEEE Internet of Things
  Journal}, vol.~8, no.~2, pp.~901--915, 2021.

\bibitem{8884111}
K.~Q. Abdelfadeel, D.~Zorbas, V.~Cionca, and D.~Pesch, ``{ $FREE$
  —Fine-Grained Scheduling for Reliable and Energy-Efficient Data Collection
  in LoRaWAN},'' {\em IEEE Internet of Things Journal}, vol.~7, no.~1,
  pp.~669--683, 2020.

\bibitem{9530755}
R.~Hamdi, E.~Baccour, A.~Erbad, M.~Qaraqe, and M.~Hamdi, ``{LoRa-RL: Deep
  Reinforcement Learning for Resource Management in Hybrid Energy LoRa Wireless
  Networks},'' {\em IEEE Internet of Things Journal}, vol.~9, no.~9,
  pp.~6458--6476, 2022.

\bibitem{10.1186/s13638-020-01783-5}
G.~Park, W.~Lee, and I.~Joe, ``{Network Resource Optimization with
  Reinforcement Learning for Low Power Wide Area Networks},'' {\em EURASIP J.
  Wirel. Commun. Netw.}, vol.~2020, sep 2020.

\bibitem{9802526}
L.-T. Tu, A.~Bradai, O.~B. Ahmed, S.~Garg, Y.~Pousset, and G.~Kaddoum,
  ``{Energy Efficiency Optimization in LoRa Networks—A Deep Learning
  Approach},'' {\em IEEE Transactions on Intelligent Transportation Systems},
  pp.~1--13, 2022.

\bibitem{schulman2017proximal}
J.~Schulman, F.~Wolski, P.~Dhariwal, A.~Radford, and O.~Klimov, ``{Proximal
  policy optimization algorithms},'' {\em arXiv preprint arXiv:1707.06347},
  2017.

\bibitem{reynders2017power}
B.~Reynders, W.~Meert, and S.~Pollin, ``Power and spreading factor control in
  low power wide area networks,'' in {\em 2017 IEEE International Conference on
  Communications (ICC)}, pp.~1--6, IEEE, 2017.

\end{thebibliography}

\end{document}